\documentclass[
  journal=BJPS,
  manuscript=article-type,
  year=2025,
  volume=XX,
]{cup-journal}

\usepackage{amsmath}
\usepackage[nopatch]{microtype}
\usepackage{booktabs}
\usepackage[backend=biber,natbib=true,style=chicago-authordate-trad]{biblatex}

\usepackage{tabularx}
\usepackage{makecell}
\usepackage{hyperref}
\newcolumntype{Y}{>{\raggedright\arraybackslash}X}
\usepackage{placeins}
 \usepackage{xcolor}
 \newcommand{\textoverline}[1]{$\overline{\mbox{#1}}$}

\title{Expressing Political Identity Online:\\
Differential issue-based self-identification across the Left–Right spectrum
}

\author{Carlo Santagiustina}
\affiliation[1]{(1) INRIA, Paris, France; (2) médialab, Sciences Po, Paris, France;  (3) VSM, Ca'Foscari University, Venice, Italy}
\email[Carlo Santagiustina]{carlo.santagiustina@inria.fr}
\author{Jean-Philippe Cointet}
\affiliation{médialab, Sciences Po, Paris, France}
\author{Pedro Ramaciotti}
\affiliation[2]{(1) Complex Systems Institute, CNRS, Paris, France; (2) médialab, Sciences Po, Paris, France;  (3) Lattice CNRS, ENS-PSL \& Sorbonne Nouvelle, Paris, France}

\begin{document}
    
\begin{abstract}
Political orientations are known to be closely linked to positions on identity-related issues, whose salience and framing can vary both within and across countries. To capture patterns of issue-based self-identification and address pressing questions about identity (mis)alignment across EU countries and between Left- and Right-leaning user groups, we study Social Media profile bios from Twitter/X, where individuals present themselves. Our analysis combines Structural Topic Modeling seeded with Manifesto Project data, and Ideology Scaling calibrated with political expert survey data. The proposed approach allows us to study how more than 1.2 million individuals from eight EU countries self-identify with political issues across the Left-Right ideological spectrum.
We find that issues such as \textit{democracy}, \textit{national way of life}, and \textit{decentralization} emerge as the most divisive at the EU level. Other issues, such as \textit{environmentalism}, \textit{equality}, and \textit{freedom \& human rights}, among others, strongly differentiate how Left- and Right-leaning individuals communicate their identity online at the EU level.
Interestingly, we also find that user groups on the Left side of the political spectrum express their identities more similarly across European countries than those on the Right. This pattern is particularly evident for identity-issues related to \textit{national way of life}, suggesting that the growing international coordination among conservative parties is not (yet) accompanied by an alignment of Right-leaning users' communicated identity across the targeted EU countries. Finally, we show that our Left-Right identity similarity metrics are inversely related to country-specific perceptions of political conflict, as measured by the PEW survey. Our findings provide a major contribution to comparative political identity studies, enabling a qualitative and quantitative characterization of the EU identity puzzle and offering empirically grounded insights for better understanding identity-related mobilization on Social Media, both within and across countries.
\end{abstract}
\newpage
\section{Introduction}
Research on issue-based political identification \citep{Huber1989values,Kluver2016} suggests that expressed identities can shape political mobilization, voting, and also European integration \citep{DeVries2007sleeping,Hooghe2002does}, with issues such as national identity and patriotism \citep{Flanagan2003newpolitics} multiculturalism and diversity \citep{Mair2007}, moral traditionalism  \citep{Knutsen2004,DeKoster2007cultural}, and environmentalism \citep{GreenPedersen2019reshaping} acting as salient political identity dimensions. These dimensions are thought to distinguish groups and individuals according to their political orientations. However, recent studies \citep{Steiner2024,Hooghe2002does,Langsaether2019} have highlighted the complexity of political identity puzzles, showing that the issues with which individuals and groups identify across the ideological spectrum may vary over time, across cohorts, and national contexts. Moreover, a recent work by \citet{clemente2024online} shows that political identities are becoming increasingly salient on Social Media platforms, like X/Twitter, and that users’ willingness to express them has also grown over the past decade, in a way that is not merely due to a changing composition of users on this platform.\\

To address this puzzle, we operationalise a new approach that allows us to study issue-based identity expression of Left vs. Right user groups within and across countries. Thanks to the proposed work, we are able to investigate how Social Media users from eight European Union (EU) countries integrate political issues into their online identities through their profile bios, across the Left-Right ideological spectrum. These bios, as concise expressions of identity cues \citep{Taylor2023identity, Hopkins2024rise}, offer a unique lens for understanding online political self-representation and authenticating narratives \citep{reddi2023identity}, serving as reflections of users’ political stances on issues that characterize the societal and political debate at both the national and EU level. As shown by \citet{Taylor2023identity} these cues are highly relevant, as online content evaluation is ``\textit{mediated by identity cues}'' and because these cues also influence online political discourse by shaping patterns of online content diffusion.\\

Despite the growing centrality of the aforementioned literature, which examines identity dynamics through Social Media data, current research lacks comparative studies that bridge diverse national contexts while proposing quantitative measurements and enabling rigorous cross-group inference and analysis of issue-based political self-identification alignment in users' expressed identities, for example in Social Media bios. As detailed in the conceptual framework described and visualized in section \ref{subsec:salience_framing}, by alignment, we here refer to the degree of salience (i.e., coverage) given to expressed identity-related issues from a given political orientation and/or country group, and to the way these issues are similarly or differently framed within or across these groups, at both the national and EU level.\\

The present work addresses the aforementioned gap by proposing a methodology capable of answering, and providing empirical grounding for, the following research questions: (i) which issues are made salient within user biographies across different political orientation groups and EU countries? (ii) how the framing of these issues varies across countries and by political orientation? and (iii) to what extent issue framing by ideological orientation is consistent or heterogeneous within and across national contexts?\\

The study is operationalized through a three-layer approach (detailed in section \ref{sec:methods} and in the sections S.2-4 of the Supplement) combining Structural Topic Modeling  \citep{Roberts2014structural}, ideology scaling of interactional trace data on Social Media calibrated with survey data for cross-country comparability \citep{Ramaciotti2024dataset}, and issues seeding from the Manifesto Project annotated corpus \citep{Merz2016}.  \\

This work differs from other recent studies \citep{Hopkins2024rise,Essig2024partisan,GonzalezBailon2023asymmetric,Taylor2023identity,kossowska2023internet, clemente2024online} along (at least) two important dimensions, which characterise our contribution: the data strategy and the methodology used to answer our research questions. Regarding data, our work focuses on a subject ---the expression of identity through bios across countries and ideological groups--- that, despite the increasing abundance of Social Media studies, remains unexplored in political communication and comparative politics research. Methodologically, the article adopts a radically different research design from existing works (for example, with respect to \cite{clemente2024online}) by leveraging seeded STM for issue inference and modeling, MP follower networks for ideology scaling, and the Manifesto Project ontology for aligning bio contents to issues that are meaningful to researchers in the political sciences. Together, these three methodological ingredients provide exactly what is required to address the aforementioned research questions, which are currently unanswered.\\

This work more broadly contributes to the research and debate on issue-based political self-identification \citep{Huber1989values,Kluver2016} in three ways: First, this is the first study to jointly model ideology-driven issue framing and issue salience across groups, using public user identity cues extracted from Social Media bios. The proposed approach allows us to explore political identity signals that resonate within specific ideological groups and/or countries, shedding light on how discourse in Social Media bios reveals identity divides between the Left and the Right, within and across EU countries. Second, by enabling comparisons of how identity-related issues are covered and framed by different groups, we differentiate high-heterogeneity, high-salience issues ---potential focal points of societal tension--- from low-heterogeneity, high-salience issues, which act as connective threads across online ideological divides. Finally, by relating Left-Right issue-based similarity to PEW's survey-based measures of political conflict at the country level, we show that our approach captures aspects of perceived political fragmentation and conflict, which, thanks to our method, can be inferred from Social Media data at the desired geographic and temporal granularity.\\

The remainder of the article is structured as follows: the Background and Gaps section (\ref{sec:background}) situates the study within existing literature and theories of political identity discourse, highlighting the contributions and limitations of previous works and approaches that seek to study online identity expression and issue-based political identification using data from Social Media. This section also outlines the conceptual foundations of the research, along with the hypotheses derived from the existing literature. The Methods \& Data section (\ref{sec:methods}) describes the approach, operationalization strategy, datasets, and methods employed. It is complemented by an extensive Supplement, which contains all methodological details and robustness checks that, due to space constraints, could not be included in this manuscript. The Results section (\ref{sec:results})  presents the key findings and their relevance, followed by a Conclusion that discusses the broader significance of our approach for understanding the contemporary political identity landscape, within and beyond the European Union. \\

\section{Background and Gaps: Political identity, issue-based identification, and identity expression in the EU}\label{sec:background}
\subsection{Identity}

Identity encompasses both uniqueness and sameness \citep{Martin1995}, shaping how individuals and groups see themselves and others \citep{Hetherington1998expressions}, form representations of each other, and perform based on the latter \citep{dervin2012cultural}. Identities emerge through the constant interplay (and narrativization) of the preferences, expectations, pressures, and performative acts of individuals \citep{stets2009identity,Jenkins2014social}. They evolve within feedback loops of anticipation and signalling behaviour \citep{cast2003identities}, through which people interact and coordinate on who they are and aspire to be \citep{Smaldino2022}, asserting their authenticity as individuals and groups \citep{archer2008younger,Schmader2018} and being recognized ---or not--- for it by others. Individuals and groups can appropriate and embody multiple identities \citep{James1890principles}, shifting between them depending on context and opportunity \citep{Yavuz2004opportunity}. Each context can elicit different aspects of identity, shaped by the implicit rules and roles that govern those interaction settings \citep{burke1981link}. 

\subsection{Political identity and issue-based self-identification}

Multiple strands of research in political studies have incorporated the notion of identity \citep{Huddy2001social,SimonKlandermans2001,Klandermans2014identity}. By focusing on identity, scholars aim to understand the underpinnings of political engagement and the ways in which identity dynamics intersect with ideology, social structures, voting, and policy making.  Scholars have explored, among others, how identity-related factors influence public opinion \citep{Hooghe2005calculation,Helbling2016mobilisation}, voting behaviour \citep{pattie1999partisanship}, political polarization \citep{WestIyengar2022}, and policy preferences \citep{Klar2013}. Political identities are important because they can affect collective coordination and deliberation \citep{Somer2018}, or exacerbate ingroup favouritism \citep{Turner1979}. Fragmented social identities can also intensify ingroup-outgroup conflicts \citep{bonomi2021identity}.\\

The structuring of political competition and issue-based mobilization across European politics is traditionally analysed through the lens of positioning on the Left-Right ideological spectrum \citep{Kluver2016,DeVries2007sleeping}. This body of research examines how identities shape attitudes and behaviours \citep{bonikowski2017nationhood}, with issues such as national identity and patriotism \citep{HuddyKhatib2007patriotism} and multiculturalism \citep{Spencer1994,Tempelman1999} dominating the research agenda. These issues are relevant because their coverage and prominence influence where credit or blame is assigned \citep{aldrich2014blame} and which policies on the agenda are prioritized by different camps.  \citet{Huber1989values} and \citet{Steiner2024} emphasize the role of national-context dynamics and cohort effects in shaping Left-Right issue-based alignment and prioritization, suggesting that issue-based political identification is a time-varying and context-sensitive phenomenon \citep{Hooghe2002does}.

\subsection{Expressing identity online}
Identity dynamics are embedded within socio-technical systems and their processes, being shaped and constrained by communication supports and expression spaces through which behaviours and interactions materialize \citep{baym2015personal,Slater2007}. Affordances provided by digital platforms and media \citep{khazraee2018digitally} can cast, frame, bind, and fix identity expression. Therefore, identity should never be considered a medium-agnostic phenomenon as it unfolds within a social interaction web, constrained by affordances, power dynamics, norms, and contexts that augment or diminish communicable aspects of identity, shaping their relevance, implications, and salience \citep{McKenna2005social,Yoder2020phans}.\\

Online platforms are increasingly central to the expression of identity, and to the circulation of political identity narratives \citep{Luders2022,Hopkins2024rise}. For example, a recent work by \citet{Hopkins2022news} shows that headlines with references to identity are more likely to be clicked on and go viral on Social Media. 
According to psychologists, Social Media provide to their users affordances to satisfy their psychological ``\textit{self-identity needs of coming to know the self, expressing self-identity, and maintaining continuity of self identity}'' \citep{karahanna2018needs}, by allowing them to express their positioning in relation to issues that are relevant to them, be it by engaging in conversations by posting opinions, sharing content via retweets, or following public figures who reflect their ideological leanings \citep{Taylor2023identity,ramaciotti2022embedding,ramaciotti2022inferring,Wojcieszak2022most}.   Among the different modes of participation, one of the most salient affordance of political identity expression lies in users’ bios or profile pages \citep{Rogers2021}. These short self-curated descriptions offer a snapshot of how individuals want to be seen by others \citep{MaizArevalo2024}. \\
Although a bio is concise, it encapsulates the core aspects of one’s political beliefs, affiliations, positions, and values \citep{Essig2024partisan}, as well as family roles,  hobbies, or even pictorial symbols such as flags and emojis \citep{luxmoore2023pl}. 
As suggested by \citet{bail2022breaking}, the online signalling of identity can, among others, amplify identity-based societal divisions, making them more salient.\\

Even when online identity expression serves as an authentic extension of offline life, users face the task of deciding ``\textit{which facets of their identities to emphasise and which to mask}'' \citep{Weinstein2014}. By selecting aspects of themselves to share publicly, users engage strategically in self-identification and authenticating narratives \citep{reddi2023identity}. This shapes political dialogue in online spaces, as sharing identity cues can influence interactions with other users \citep{Taylor2023identity}, attracting those who identify with the signalled identity traits and repelling those who do not.

\subsection{Identity salience, framing and issue-based political conflicts}\label{subsec:salience_framing}

\begin{figure}[!h]
    \centering
    \includegraphics[width=1\linewidth]{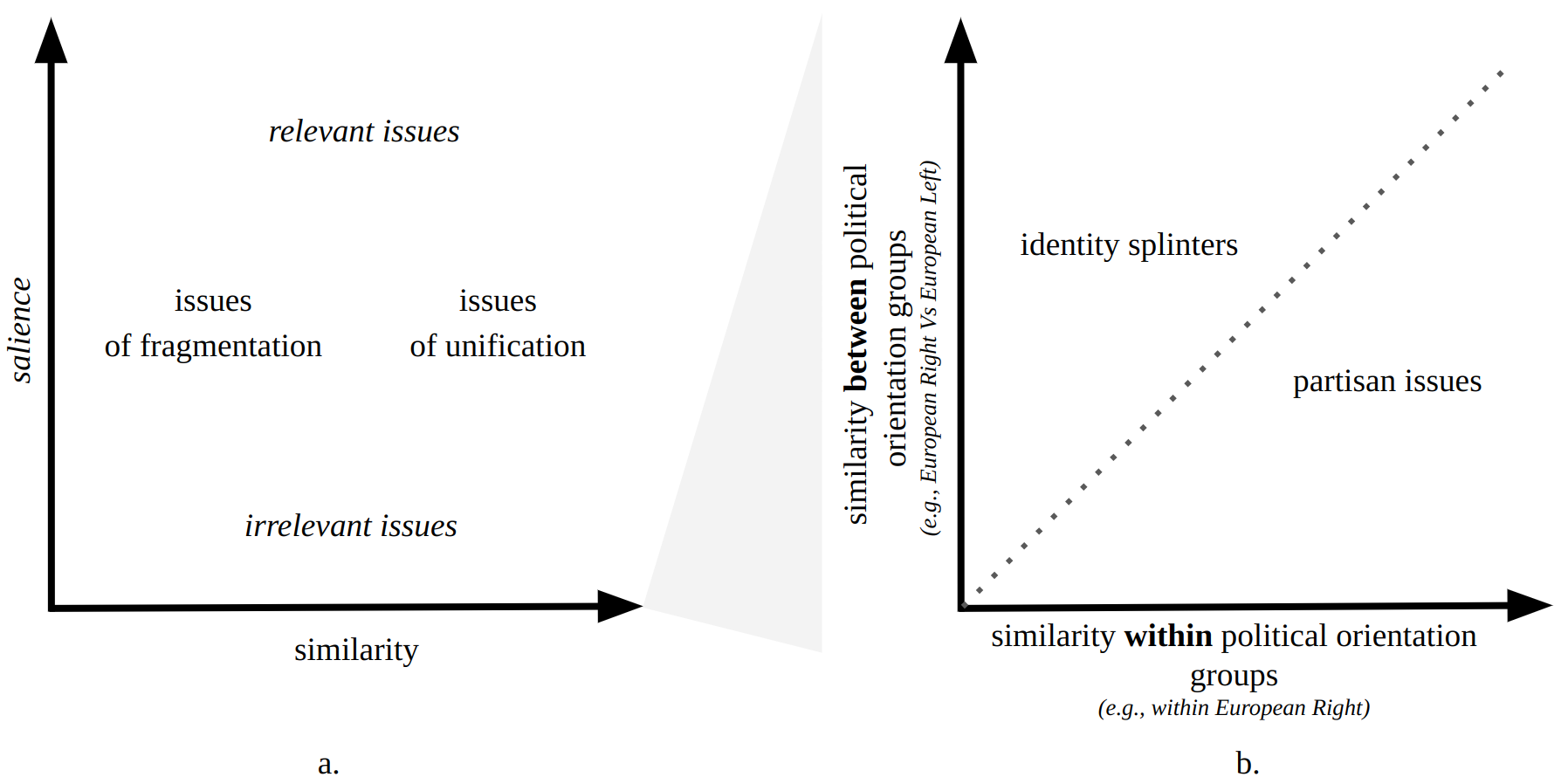}
    \caption{a. Issue types as a function of  salience and similarity across groups; b. Issue types as a function of similarity within and between political orientation groups, across countries.}
    \label{fig:theory}
\end{figure}

\citet{Eidson2017identification} describe identity framings as the qualitative base of social action, and claim that these frames, and their alignment, contribute to ``\textit{the production of relations of likeness and difference, complementarity and conflict}''. By analysing how issues are framed across different subjects and groups, one can explore perceptions of similarity both within and between groups. The likelihood that an issue is invoked by a member of a specific group \citep{Stryker1968}, also called issue salience, is a key dimension of issue-based identification, as it mediates, within ideological groups, issue visibility and alignment, and can influence the emergence of homophilic social interactions \citep{McPherson2001}. Moreover, issues that are the most salient and heterogeneous across political orientation groups, have the potential to act as flashpoints for identity-based conflicts, amplifying perceptions of societal tensions and political divisions \citep{Volkan2019}. On the other hand, issues that exhibit high salience and low heterogeneity (i.e., high similarity across groups) might foster cohesion and shared identity, reducing perceptions of conflict across groups, and creating common ground for a collective, multi-partisan identity. Whereas, issues that show low salience and high heterogeneity, can be considered hidden identity fault lines \citep{Maynard2015}. They may not impact online social interactions until activated by specific events, at which point their latent divisive potential can emerge.\\

Issue framing \citep{benford2000framing} heterogeneity can be analysed along two other perspectives.  The first examines variations in the framing of a specific issue within the Right across different countries, or, within the Left across different countries. The second considers the degree of issue framing heterogeneity between the Left and the Right, across countries. This distinction allows for the identification of two cross-country heterogeneity patterns: (i) Issues that exhibit greater heterogeneity (i.e., lower similarity) between political orientation groups than within them can be categorized as ``\textit{partisan issues}'' \citep{Peterson2018}. This pattern reflects the unifying influence of political orientation, where shared ideological alignment overrides country differences. (ii) Issues where heterogeneity within political orientation groups exceeds heterogeneity between them can be classified as ``\textit{identity splinters}'' \citep{puri2008encountering}.\\

The aforementioned theory-driven distinctions can be visually summarized using two-dimensional graphs, as shown in Figure~\ref{fig:theory}. For a more in depth explanation of this framework and related conceptual distinctions, we refer the reader to section S.1 in the Supplement.

\subsection{Alignment and differences in political identities across countries: The case of the EU}
The EU's multi-level governance structure, and possible tensions arising from the interaction between national and European politics, make it a relevant case for analysis through the proposed approach. As a supranational entity, the EU encompasses a diverse set of identity-related issues that structure national politics. In European countries, greater attention to issues related to (H1) \textit{national way of life}, \textit{traditional morality}, and \textit{law and order} is believed to be associated to conservative positions and Right-leaning political orientations \citep{Flanagan2003newpolitics,Knutsen2004,DeKoster2007cultural}, while (H2) \textit{multiculturalism} and \textit{environmentalism} is instead generally associated to progressive positions and Left-leaning political orientations \citep{Mair2007,GreenPedersen2019reshaping,Steiner2024}. \\

EU countries have distinct political histories and collective memories that can give rise to varying political identities \citep{verovvsek2016collective,Wang2017memory}. We therefore expect (H3) some of the above-mentioned issues, in particular \textit{national way of life}, to be framed differently across the political spectrum and within varying national contexts \citep{hameleers2021effects}. For example, the Italian Right is expected to frame \textit{national way of life} also in relation to Christian heritage and traditional family values, while the French Right through secularism and Republican ideals. Since issues do not exist in isolation, as they are linked to values and preferences, through the EU integration, individuals increasingly relate to movements and identities that go beyond their nation-state, looking to other countries and peoples. This gives rise to multi-level identity layers that can be studied and related to each other through the proposed approach, to study differentiated European integration \citep{Moland2024}, and identity-related conflicts.

\section{Methods \& Data}\label{sec:methods}

\begin{figure}
    \centering
    \includegraphics[width=1\linewidth]{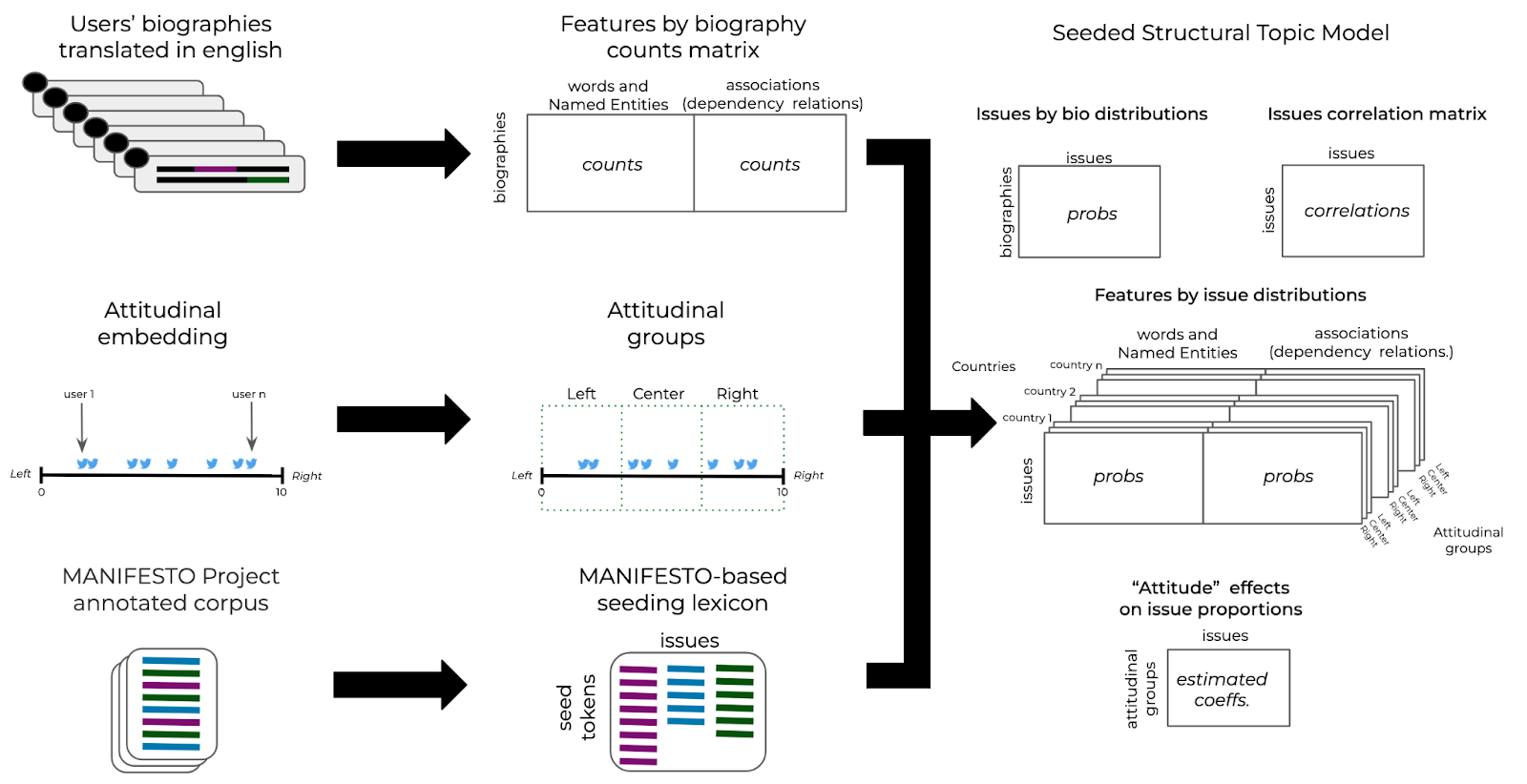}
    \caption{Overview of the proposed approach}
    \label{fig:pipeline}
\end{figure}

We operationalise our study through a three-layer approach that integrates seeded Structural Topic Modeling \citep{Roberts2014structural}, Left-Right ideology scaling method from \citet{ramaciotti2022inferring}, and the Manifesto Project’s annotated corpus \citep{Merz2016} and its ontology of political issues, which is here used to build a STM seeding lexicon. By combining these three methods, we can map and examine Left–Right and cross EU-country variations in self-representation within Social Media bios. 
The proposed approach is summarized in Figure~\ref{fig:pipeline}.

Social Media platforms, like X, are virtual spaces where identities are expressed, contested and (re)negotiated \citep{hanckel2019s, marwick2013online}. As Social Media platform affordances and norms encourage users to curate short bios that are easily digestible, it facilitates expressed identity comparability. We here focus on X, as, despite recent crowding-out movements that followed Elon Musk’s takeover \citep{claesson2024twitter}, until the 2024 US presidential elections, it has been the Social Media platform of choice for a large and active population of politicians and elected officials \citep{VanVliet2020}, journalists \citep{Revers2014,Molyneux2022}, and experts \citep{Zhang2024} on a global scale.
Since we are interested in analysing differential issue-based self-identification across political orientations and countries, we need to select a population of individuals that lend themselves to spatial political representation (i.e., displaying sufficient political sophistication; see \cite{luskin1990explaining}). MPs on X constitute a population that plays a consistent role in the online political spheres across countries \citep{VanVliet2020,ramaciotti2022inferring}. Therefore, national parliaments provide a standardized comparative layer that allows us to delimit the targeted population for each country. \\

Our study builds on the dataset of Members of Parliament (MPs) and their followers on X by \citet{Ramaciotti2024dataset}, which includes the five most populous countries in the EU  (Germany, France, Italy, Spain, Poland) as well as the Netherlands, Belgium and Slovenia, all of which have a high rate of MPs with active X accounts. This dataset was seeded with a manual identification and annotation of accounts of MPs during January, February, and March 2023, and their followers, collected during March 2023 using the platform’s API. 
Starting from the MPs on X from the selected countries, we consider their followers as our populations of study. MP followers on X have been shown to be exploitable with ideology scaling methods to infer their positions in Left-Right scales, in several settings \citep{barbera2015tweeting} and validated using voting records, affiliation to social movements, and media diets \citep{barbera2015birds}. 
MPs-follower networks have been shown to be spatializable in ideology dimensions using scaling methods in several contexts \citep{barbera2015birds,barbera2015tweeting}. The spatialization we use includes a Left-Right scaling of MPs and their followers, calibrated using the 2023 Chapel Expert Survey \citep{jolly2022chapel}. This dataset provides comparable political spatial references. The Left-Right dimension we use captures MP followers’ political orientation along the Left-Right spectrum, and is based on the CHES Left-Right (LR) scale, which ranges from 0 to 10. To simplify this dimension for inference, we discretized it into three equal-sized categories: Left (0–3.33), Center (3.34–6.65), and Right (6.66–10). This discrete variable is used as covariate within our Structural Topic Model, influencing both issues’ proportions and issues’ contents, in combination with the user's country covariate\footnote{The authors acknowledge that partitioning the political Left-Right spectrum into three groups per country is a simplification. However, this approach is necessary for the Structural Topic Model (STM) library \citep{Roberts2019stm}, which requires categorical (rather than continuous) variables for content covariates. Moreover, the decision to use three  categories for each country is motivated by the need to manage the number of categories in the STM estimation, where it directly affects topic content and topic propensity estimates. Specifically, this partitioning results in 24 levels/groups (8 countries × 3 political orientations), helping to keep the overall complexity of the STM and of our results manageable, despite the richness of our study. It also allows for more direct comparisons between the Left and the Right, without generating an excessive number of combinations, which aligns with the objectives of this work. We explored the possibility of dividing the political space into five or more groups per country, but this resulted in having too few observations per group in smaller countries, such as Slovenia, thereby undermining the reliability of our estimates. For these reasons, we consistently partition the political spectrum into three user groups for each country. In future work, for a single country with enough observations (e.g., France and Germany) we envisage to split the Left–Right spectrum more finely to compare how moderate and radical left- and right-wing users express identity on Social Media, and how expressed identity has changed across time.}.  For details and robustness checks, please refer to sections S.2 and S.4 in the Supplement.\\

To prepare the bios of MP followers for the analysis, we filter out all users with empty or very short bios (i.e., >2 features). Our filtered dataset contains 1 201 600 observations (i.e., X user bios in English\footnote{To make the bios comparable across EU languages, we use English-translated versions. Bios have been translated to English using the M2M100 multilingual encoder-decoder 1.2B parameters model by Facebook. Available here: \href{https://huggingface.co/facebook/m2m100_1.2B}{https://huggingface.co/facebook/m2m100\_1.2B} \\
A random sample of 250 observations was manually checked by the authors, and compared to Google translated versions, to evaluate the quality of the automated translation. Except for the arbitrary addition or removal of emojis, which are not considered in this work, and a few minor translation errors, no artifacts were observed.%
}) from the eight targeted countries and related metadata. The core of our approach relies on a Structural Topic Model of Social Media users' biographies, estimated using both words and dependency relations, which is seeded with a lexicon (and an ontology of issues) derived from the Manifesto Project \citep{Merz2016}, which is the ontological layer of our method. \\

The Manifesto Project is a resource that provides annotated texts from political party manifestos, also translated in English, offering a theoretically grounded source of phrases with labels related to political issues from a variety of domains. 
By seeding the STM with a lexicon based on Manifesto Project phrases annotated with issue labels, we ensure that the inferred issues are closely aligned with a theoretically-grounded and well established ontology of political issues. Issues in the seeding lexicon are shown in Figure~\ref{fig:salience}. For details about the seeding process, the lexicon construction and content, please refer to section S.3 in the Supplement.
This lexicon is hence used to seed our STM, guiding the model to associate relevant words and word-associations with theoretically grounded issues from the Manifesto Project\footnote{We also add a residual issue, with a non-informative prior, to capture themes that are not captured by the the seeding lexicon, this also allows other seeded issues to better specialize on the targeted issues (i.e.,  those from the Manifiesto Project ontology).}. \\

By incorporating the country and Left-Right dimensions as covariates in our model, STM can account for variations in how users with different political orientations express their identities across issues and EU countries. For example, one might investigate how environmentalism is framed differently by individuals on opposite ends of the political spectrum\footnote{For related results, refer to Figure~S.11 and Table~S.4 in the Supplement}, focusing on the distinct words and associations each side employs. For instance, showing that the Left stresses urgent policy action on the climate crisis, while the Right emphasizes animal welfare.\\

The LR (Left, Center, Right) and country (Belgium, France, Germany, Italy, the Netherlands, Poland, Slovenia, Spain) categorical variables, representing respectively the ideological positioning of users from the ideology scaling and the country of the MPs they follow, are combined and included as covariates affecting both issue proportions and issue content. This modelling approach enables a systematic exploration of the framing of identity-related issues in Social Media profile bios, allowing for a deeper understanding of the interaction between political orientations and issue-based identity narratives within and across EU countries. For details on the STM specification, inference strategy and robustness checks, please refer to section S.4 of the Supplement.

\section{Results}\label{sec:results}
In this section we presents the main findings of our study, which are devised into three subsections. First, in Subsection \ref{subsec:issue_salience}, we examine issue salience across EU countries and political orientations, identifying issues that vary in coverage in users' bio by ideological leaning and country. Second, in Subsection  \ref{subsec:framing}, we analyse issue framing content similarity across EU countries and political orientations, comparing how distinct ideological groups frame more or less similarly a given issue, at the EU level. Results reveal, among others, that Left-leaning users display greater cross-country alignment in issue framing than Right-leaning ones. Finally, in Subsection  \ref{subsec:framing_country}, we assess within-country Left–Right issue framing similarity, showing that our metrics inversely correlate with external measures of perceived political conflict. Showing that people from countries with higher Left–Right identity alignment (i.e., similarity), such as Belgium and Slovenia, tend to also report lower perceived political conflict in PEW surveys, while those with lower similarity (e.g., Germany, Spain) exhibit higher perceived political conflict. 
The results that follow allow us to map the contours of political identity expression across Europe, revealing how ideological and national factors jointly shape the online communication and co-construction of political identities.
\subsection{Issue salience across EU countries and political orientations}\label{subsec:issue_salience}
To assess the salience of the different issues mentioned in users' bios, we compute mean issue propensity scores by group, where each group of individuals corresponds to a specific country and ideological orientation combination. 
\begin{figure}[!h]
    \centering
    \includegraphics[width=1\linewidth]{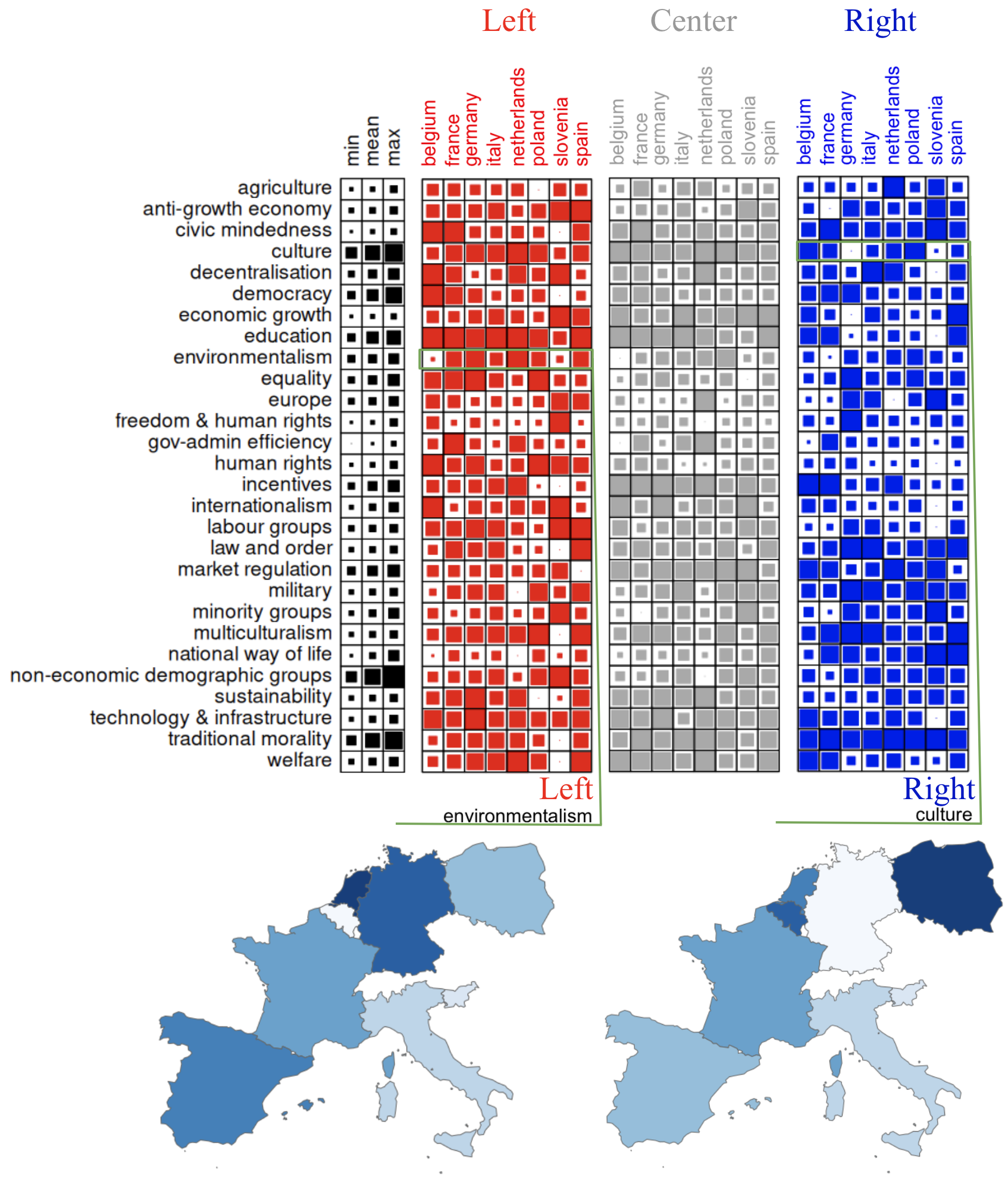}
    \caption{Average issue propensity scores by country and political orientation. The first three columns on the left display the minimum, mean, and maximum issue propensity scores by group. The size of each square represents an issue’s propensity score, defined as the average probability of an issue appearing in individuals' biographies within each political orientation and country group. Excluding the first three columns, scores are normalized within a zero-one range by row to facilitate within-issue comparisons across groups. For statistical significance tests refer to Table~S.3 in the Supplement.}
    \label{fig:salience}
\end{figure}
In Figure~\ref{fig:salience}, we highlight differences in average issue propensity score as a function of individuals’ country and political orientation. The first three columns on the left allow us to compare the minimum, mean, and maximum issue score by group. To facilitate within-issue comparisons by group. Excluding the first three columns, issue propensity scores by group in Figure~\ref{fig:salience} have been normalized by row.  \\

Figure~\ref{fig:salience} and Table~S.3 in the Supplement (which also displays statistical  significance tests), show that some issues, like \textit{economic growth}, \textit{technology \& infrastructure}, \textit{gov-admin efficiency}, \textit{civic mindedness}, \textit{market regulation}, \textit{anti-growth economy}, and \textit{welfare} exhibit rather consistent levels of prevalence across the Left, Center, and Right, and also across countries, suggesting a shared baseline interest/concern for these issues within the EU, which transcends ideological boundaries within and across countries.\\
In contrast, other issues, like \textit{culture}, \textit{education}, \textit{democracy}, \textit{incentives}, \textit{europe}, \textit{decentralisation} and \textit{traditional morality} show significant variations in coverage across political orientation groups and among EU countries. For instance, \textit{education} is notably less emphasized in Slovenia and Poland, compared to other countries, who generally give \textit{education} a relatively higher coverage in their bios. Also, on the Right side of the political spectrum \textit{education} receives comparatively low salience when contrasted with the Left, highlighting a notable difference in how \textit{education} features as part of group’s identity communication. This finding is consistent across most countries, underscoring a cross-national trend in which education-related identity cues are, on average, more relevant to Left-leaning individuals. For other issues, like \textit{democracy}, differences between political orientation and country groups are also very pronounced. While \textit{democracy} tends to be more prominent overall in Left-oriented users groups compared to Center- and Right-oriented ones, there are significant country-specific differences among countries. For instance, Left-oriented user groups in countries like Slovenia show significantly lower scores for alluding to the issue of \textit{democracy}. \\

Figure~\ref{fig:salience} shows that some issues, like \textit{national way of life}, \textit{environmentalism}, \textit{equality}, \textit{internationalism}, and \textit{technology \& infrastructure}, and \textit{welfare} present large variations across countries. For example, users from Spain, Poland and Italy tend to dedicate more space to the \textit{national way of life} issue. This emphasis forms a distinctive dimension of how individuals from these countries communicate their identity on Social Media, highlighting issues related to national pride, patriotism and migration as central elements of their bio.  \textit{Environmentalism} also exhibits significant country-specific variations, with nations like the Netherlands and Germany demonstrating higher issue propensity scores. Interestingly, the Netherlands and Germany, which exhibit the two higher scores for discussing \textit{environmentalism} in bios, also ranked among the top 15 in the World Economic Forum’s 2024 Environmental Performance Index \citep{block2024epi}, suggesting that issue salience may capture the diffusion of (national and/or political group norms) into online self‑presentation; in other words, reflecting, among others, political discourse and priorities at the national level. 
For further details, refer to section S.5 in the Supplement.\\
\FloatBarrier
\subsection{Issue framing similarity across countries and political orientations}\label{subsec:framing}

\begin{figure}
    \centering
    \includegraphics[width=1\linewidth]{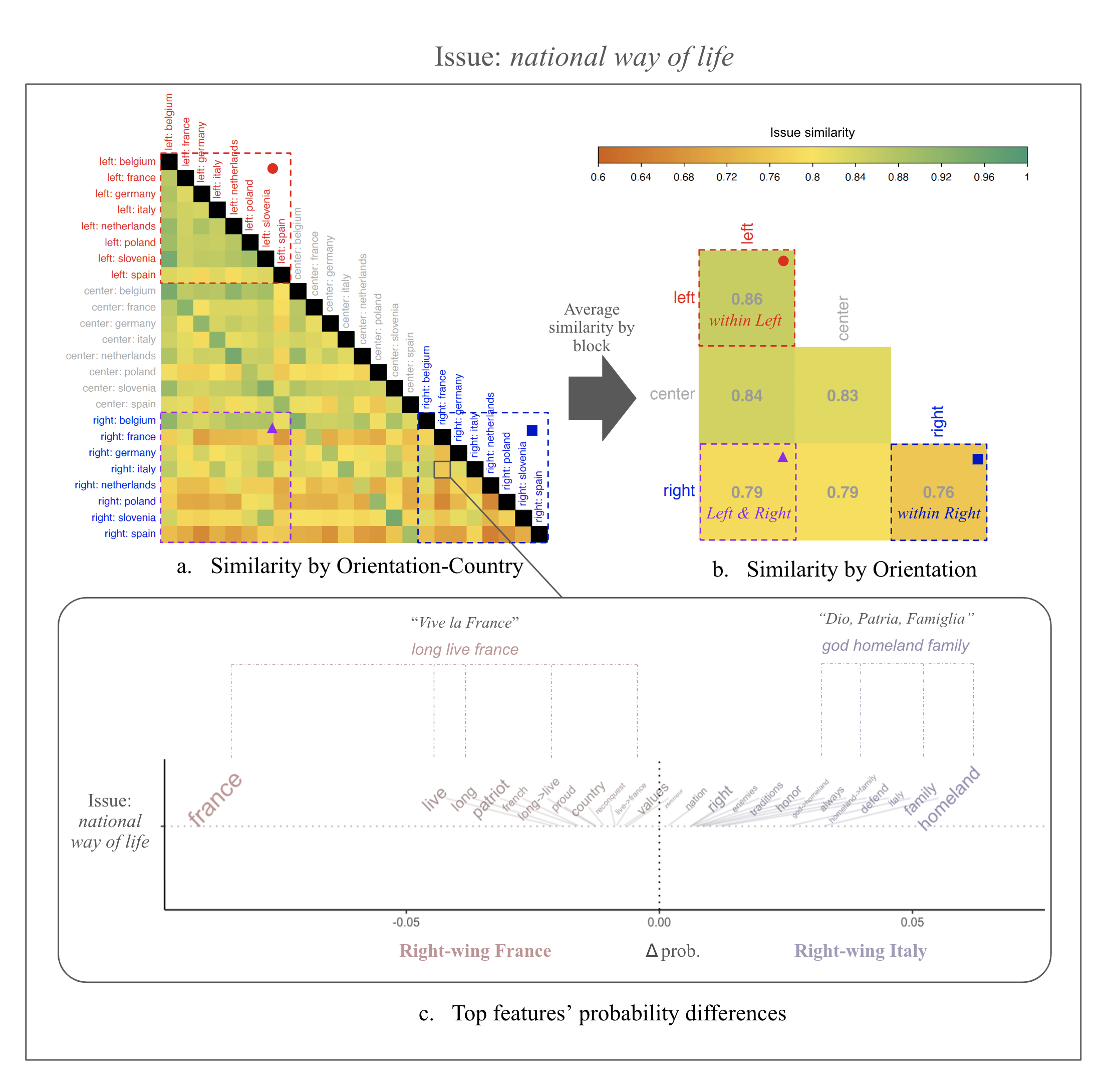}
    \caption{Issue framing similarity for \textit{national way of life}. (a) Color-encoded matrix displaying pairwise similarities (1 - JS divergence) between issue contents (i.e., features distributions) by Orientation-Country group (pol.pos.g: country.j) for the selected issue. (b) Color-encoded matrix showing average issue similarities by block, both within and between political orientation groups (Left, Center, and Right) for all EU countries in our dataset. (c) Top feature probability differences between the right-wing groups in France and Italy. The size of each feature is proportional to its average probability for the two groups being compared. The color of the features and their position on the x-axis represent differences in probabilities between the Right and the Left. For any feature  $i$, probability differences are computed as follows: \\
$\Delta prob_i=prob(\text{feat}_i|\text{right:italy},\text{national way of life})-prob(\text{feat}_i|\text{right:france},\text{national way of life})$\\
Annotations in French and English have been manually added to highlight that, among the top Right-wing words and associations for the two countries for the \textit{national way of life} issue, we find many features related to historic patriotic mottos. }

    \label{fig:similarity}
\end{figure}

To assess, for a given issue, how similar the words and associations used by different user groups are, we compute the Jensen-Shannon (JS) divergence between features-by-issue distributions. Using log base 2, JS divergence is bounded between 0 (identical distributions) and 1 (very dissimilar distributions). These distributions inferred through STM represent the probabilities of features (i.e., words and associations) observed in users' bios by issue,  political orientation, and users’ country group. To measure the heterogeneity within the Left-wing and the Right-wing across our sample of EU countries ---and to quantify how the two sides of the political spectrum differ--- we calculate the average JS divergence by issue, and by political orientation\footnote{As a robustness check, we also calculated the Generalized Jensen-Shannon Divergence (GJSD), using the philentropy R library \citep{drost2018philentropy}, which allows us to measure the divergence between more than two probability distributions. The results and findings obtained are similar to those derived from the non-generalized version presented in the main paper, indicating that related findings are not sensitive to the averaging method used to quantify distributional differences within and across groups.}. As illustrated in Figure \ref{fig:theory}, this can be done within political groups (e.g., within Left and within Right) and between groups (e.g., Left \& Right), the latter representing the average pairwise distance between Left-leaning and Right-leaning user groups across the eight targeted countries. 
In Figure~\ref{fig:similarity} we show the similarity (1-JS divergence) between groups for the issue \textit{national way of life}. Figure~~\ref{fig:similarity}.a. reveals, among others, that the way this issue is framed by Right-leaning users in Italy is more similar to that of Right-leaning users in Germany (0.82), compared to Right-leaning users in France (0.76) or Spain (0.78). 
Figure~\ref{fig:similarity}.b., shows within and across political orientation groups similarity averages among the eight investigated countries. It reveals that, for \textit{national way of life}, there appears to be more similarity between EU-countries' identities expressed by user groups on the Left side of the political spectrum (0.86) compared to the Right (0.76). This suggests that there are structural differences in how the two sides of the political spectrum express their identities across the investigated countries, warranting further analysis if this pattern is observed only for this issue or also for others.
To assess the differences in feature probabilities for the compared groups, which drive our similarity metrics, we also investigate, through Figure~\ref{fig:similarity}.c, the differences feature probabilities for the top 15 features for the Right-wing in France \textit{vs.} the Right-wing in Italy, for the issue \textit{national way of life}. This allows us to explore which words and associations differentiate Right-wing users in these two countries for what concerns \textit{national way of life}.\\

Words and associations on the left of the plot (such as ``\textit{france}'', ``\textit{long->live}'', ``\textit{patriot}'', ``\textit{proud}'') characterize, for the selected issue, the bios of French Right-wing users, while those on the right side of the plot (such as ``\textit{homeland}'', ``\textit{god->homeland}'', ``\textit{family}'', ``\textit{honor}'', ``\textit{defend}'') characterize the Italian Right-wing group. Terms near the center (like ``\textit{nation}'') are about equally prevalent in the bios of both groups for this issue.  These differences suggests users from the French Right tend to emphasize civic pride and national identity, expressed through republican patriotic language, whereas users from the Italian Right highlight more traditionalist and Christian values, where references to family, religion, and defence of the homeland appear to be more central. \\

Interestingly, many of the words and associations that differentiate the bios of these two groups for the selected issue come from historic mottos used in the two countries: ``\textit{Dio, Patria, Famiglia}''  (God, Homeland, Family) in Italy and ``\textit{Vive la France}'' (Long live France) in France. Both formulas appearing on X users' bios carry a long political genealogy and have been central to the symbolic repertoire of conservative and nationalist movements\footnote{
In the Italian case, ``\textit{Dio, Patria, Famiglia}'', first used by Giuseppe Mazzini in his book titled ``\textit{Doveri Dell'Uomo}'', became a rallying cry of the conservative Right in the early twentieth century and was institutionalized under Fascism, where it encapsulated the regime’s vision of Christian moral order, loyalty to the nation, and traditional social hierarchies. After 1945, its associations with Mussolini’s dictatorship made it a contested slogan, but it remained a reference point for far-Right postwar parties and Catholic conservatives. Today, this motto is often mobilized in defence of national identity against secularization and internationalism, and it continues to resonate among nationalist currents, also through Social Media, as a shorthand for the protection of cultural tradition and resistance to perceived threats from globalization and immigration.  
By contrast, ``\textit{Vive la France}'' has its origins in the French Revolution, when it was used as an affirmation of unity, popular sovereignty, and republican values (often in the form ``\textit{Vive la France, Vive la République}''). Over the twentieth century, it progressively became an ubiquitous patriotic exclamation, employed across the whole political spectrum. During the Second World War, it was notably invoked both by Charles de Gaulle, who concluded his speech of 22 June 1940 with the phrase ``\textit{Vive la France libre dans l'honneur et dans l'indépendance}'', and by Philippe Pétain, whose regime used the slogan ``\textit{Vive Pétain, vive la France}'' in political communication materials. Gaullist as well as conservative nationalist movements tend to invoke this motto to stress continuity with the past and to assert cultural pride in opposition to perceived external threats.}. In recent times, these patriotic mottos have been publicly re-appropriated by nationalist political elites in both countries: Marine \citet{le2012pour} titled her book ``\textit{Pour Que Vive la France}'', while Giorgia Meloni, has repetitively invoked ``\textit{Dio, Patria, Famiglia}'' as her core values in her speeches, for example during the 2022 general election \citep{corriere2022meloni}. She has also had the motto printed on banners for use at official events and rallies of her party, Fratelli d'Italia (\href{https://www.fratelli-italia.it/marcia-per-la-vita-fdi-partecipa-con-sua-delegazione-orgogliosi-di-difendere-dio-patria-e-famiglia/}{Link to Fratelli d'Italia official webpage containing a photo of the banner used during a March for Life}), and, since 2019, posted photos of herself waving a scarf emblazoned with this slogan, notably on her official Facebook profile (\href{https://www.facebook.com/giorgiameloni.paginaufficiale/posts/con-buona-pace-di-chi-vorrebbe-distruggere-la-nostra-identit%C3%A0/10157765611787645/}{Link to post Id:10157765611787645}), accompanied by the text: ``\textit{With all due respect to those who would like to destroy our identity}''.\\

The wide use of these national mottos in right-wing users' bios,  illustrates how conservative and nationalist movements in Italy and France draw on historical symbolic capital to culturally-root and legitimize their political identities. Although we cannot establish a causal link between the presence of these mottos in users' bios and their recent re-appropriation by radical Right political elites, we hypothesize that by referring to historical patriotic slogans, both leaders ---as well as their online supporters--- seek to situate their identity discourse within a lineage of nationalist mobilization, tapping into deep-seated cultural memories that, as our findings show, strongly resonate with ideologically aligned user communities on Social Media platforms. These findings also speak to literature about the propagandistic use of identity and authenticating narratives \citep{reddi2023identity}, such as the recent work by \citet{horz2024identity}, which shows how political leaders and elites can exploit their agenda-setting power to (re)design identity norms and mobilize aligned citizens.\\ 

This example, which focuses on the framing of national identity within the right side of the political spectrum in France versus Italy, illustrates one of many meaningful differences that can be identified and explained by examining similarities in identity framing across Orientation-Country groups as inferred through the proposed approach. Furthermore, it highlights how exploring the words and associations that drive cross-country and/or cross-political orientation issue framing (mis)alignment can provide qualitative insights that complement our quantitative metrics for issue salience and framing alignment. In the Supplementary materials (Section S.6), we provide an extended description and interpretation of other notable differences in issue framing across the Left-Right ideological spectrum, at the EU level.\\

\begin{figure}[!h]
    \centering
    \includegraphics[width=1\linewidth]{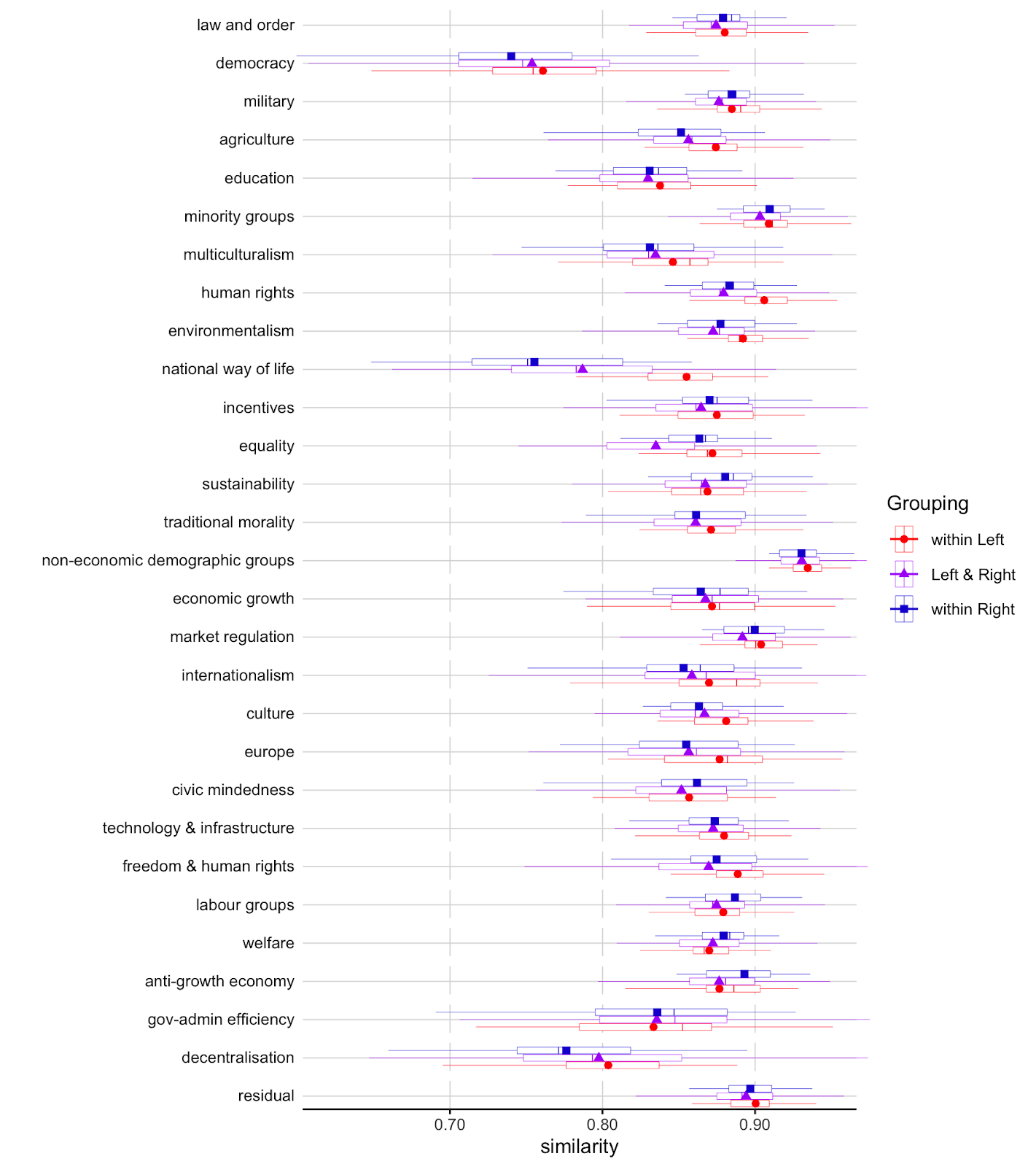}
    \caption{Shapes represent political blocks’ average similarity scores by issue and by political orientation grouping (red circles: within Left; violet triangles: Left \& Right; blue squares: within Right). Box-plots represent the dispersion of issue similarities between pairs of countries, by grouping (red boxes: within Left; violet boxes: Left \& Right; blue boxes: within Right).}
    \label{fig:distance}
\end{figure}

Going back to our broader EU-level political identity puzzle, similarities within and between political orientation groups are summarized in Figure~\ref{fig:distance}. Which shows that, like for \textit{national way of life}, for a large number of issues,  average differences in issue contents among countries are smaller for the Left (within Left) than for the Right (within Right). Table~\ref{fig:ttest}, which displays the results of t-tests for differences between the means of the two political orientation groups (within Left and within Right) considering all countries, shows that for 11 issues (\textit{national way of life}, \textit{human rights}, \textit{europe}, \textit{multiculturalism}, \textit{decentralisation}, \textit{agriculture}, \textit{freedom \& human rights}, \textit{internationalism}, \textit{culture}, \textit{democracy}, and \textit{environmentalism}), the mean similarity within the Left is significantly higher (at the 0.05 significance level) compared to that within the Right, while the opposite is true for only 4 issues (\textit{anti-growth economy}, \textit{welfare}, \textit{sustainability}, and \textit{labour groups}).
\begin{table}[!h]
\centering
\caption{Two-sided (paired) t-test results comparing the group means of issue-content similarity scores between “within Left” and “within Right” groups, for each issue, for all considered EU countries. The table reports the mean similarity for each group (“within Left” and “within Right”), the difference between group means, the t-statistic, p-value and statistical significance (indicated by asterisks). Rows are ordered by t-statistic value, from highest to lowest. Issues at the top of the table represent topics for which there is greater content alignment within the Left, while topics at the bottom represent those with greater alignment within the Right, considering all EU countries in our dataset. \\ } 
\label{tab:ttest_results}
\begin{tabular}{lrrrrrl}
  \hline
Issue & \textit{\color{red}\textoverline{within Left}} & \textit{\color{blue}\textoverline{within Right}} & difference & t-statistic & p-value & \\ 
  \hline
national way of life & 0.8550 & 0.7553 & 0.0997 & 10.3871 & 0.0000 & *** \\ 
  human rights & 0.9060 & 0.8831 & 0.0229 & 6.4932 & 0.0000 & *** \\ 
  europe & 0.8767 & 0.8548 & 0.0219 & 6.0977 & 0.0000 & *** \\ 
  multiculturalism & 0.8461 & 0.8311 & 0.0151 & 4.8770 & 0.0000 & *** \\ 
  decentralisation & 0.8037 & 0.7763 & 0.0275 & 4.4865 & 0.0001 & *** \\ 
  agriculture & 0.8743 & 0.8515 & 0.0228 & 4.2900 & 0.0002 & *** \\ 
  freedom \& human rights & 0.8886 & 0.8748 & 0.0138 & 4.0224 & 0.0004 & *** \\ 
  internationalism & 0.8698 & 0.8531 & 0.0167 & 3.7059 & 0.0010 & *** \\ 
  culture & 0.8810 & 0.8632 & 0.0178 & 3.5382 & 0.0015 & ** \\ 
  democracy & 0.7609 & 0.7401 & 0.0208 & 2.8065 & 0.0092 & ** \\ 
  environmentalism & 0.8921 & 0.8773 & 0.0147 & 2.3825 & 0.0245 & * \\ 
  technology \& infrastructure & 0.8797 & 0.8735 & 0.0062 & 1.8671 & 0.0728 &  \\ 
  non-economic demographic groups & 0.9345 & 0.9304 & 0.0040 & 1.8084 & 0.0817 &  \\ 
  education & 0.8377 & 0.8309 & 0.0068 & 1.7550 & 0.0906 &  \\ 
  traditional morality & 0.8711 & 0.8613 & 0.0098 & 1.5592 & 0.1306 &  \\ 
  market regulation & 0.9039 & 0.8998 & 0.0041 & 1.3933 & 0.1749 &  \\ 
  economic growth & 0.8717 & 0.8644 & 0.0074 & 1.3790 & 0.1792 &  \\ 
  equality & 0.8720 & 0.8634 & 0.0086 & 1.3081 & 0.2019 &  \\ 
  residual & 0.9004 & 0.8968 & 0.0036 & 1.1448 & 0.2624 &  \\ 
  incentives & 0.8749 & 0.8702 & 0.0047 & 0.7718 & 0.4469 &  \\ 
  law and order & 0.8801 & 0.8790 & 0.0011 & 0.2928 & 0.7719 &  \\ 
  military & 0.8848 & 0.8846 & 0.0001 & 0.0260 & 0.9794 &  \\ 
  minority groups & 0.9089 & 0.9095 & -0.0006 & -0.1577 & 0.8758 &  \\ 
  gov-admin efficiency & 0.8334 & 0.8359 & -0.0024 & -0.3795 & 0.7073 &  \\ 
  civic mindedness & 0.8568 & 0.8619 & -0.0051 & -0.8942 & 0.3791 &  \\ 
  labour groups & 0.8791 & 0.8868 & -0.0077 & -2.0980 & 0.0454 & * \\ 
  sustainability & 0.8688 & 0.8804 & -0.0116 & -2.1614 & 0.0397 & * \\ 
  welfare & 0.8699 & 0.8794 & -0.0095 & -2.7969 & 0.0094 & ** \\ 
  anti-growth economy & 0.8766 & 0.8930 & -0.0164 & -2.9862 & 0.0059 & ** \\ 
   \hline
\end{tabular}
Statistical significance levels: $^{*}$p < 0.05, $^{**}$p < 0.01, $^{***}$p < 0.001.
\label{fig:ttest}
\end{table}
\FloatBarrier

This reveals that expressed identity alignment across EU countries is markedly weaker among Right-wing user groups than among their Left-wing counterparts. This finding also suggests that articulating a common conservative identity framework and sustaining a cohesive Right-wing identity narrative at the EU level may represent a great challenge for the Right, w.r.t. the Left. In this respect, the evidence presented here nuances, and to some extent challenges, a recurrent assumption in recent media discourse (see, for example, article by Le Monde: \url{www.lemonde.fr/en/opinion/article/2024/06/08/how-the-far-right-envisions-europe_6674195_23.html}), that the Right, grounded in appeals to Europe's ``traditional cultural identity'', would display higher levels of identity convergence than the Left, which is often portrayed as more fragmented. Instead, our results point to a paradox with this type of discourse: while far-Right elites and parties might invoke the language of shared identity and values, in practice the online expressions of users' identity (on X/Twitter) from the Right side of the political spectrum remain greatly inflected by EU country specificities, producing a lower degree of cross-national alignment (at the EU level) than is observable among Left-wing user groups, especially for what concerns the \textit{national way of life} issue.
Finally, Figure~\ref{fig:distance} also reveals that for ten issues (\textit{agriculture}, \textit{culture}, \textit{decentralization}, \textit{democracy}, \textit{economic growth}, \textit{europe}, \textit{internationalism}, \textit{multiculturalism}, \textit{national way of life}, and \textit{non-economic demographic groups}), the average similarity among EU countries within the Right side of the political spectrum is smaller or equal than that between the Right and Left, suggesting that, at least for these issues, finding common ground for constructing and communicating a multi-partisan identity (i.e., one which should allow individuals in both sides of the political spectrum to self-identify with at the EU level) would be easier than constructing on top of them a unified Right-wing identity. \\

\begin{figure}[!h]
    \centering
    \includegraphics[width=0.85\linewidth]{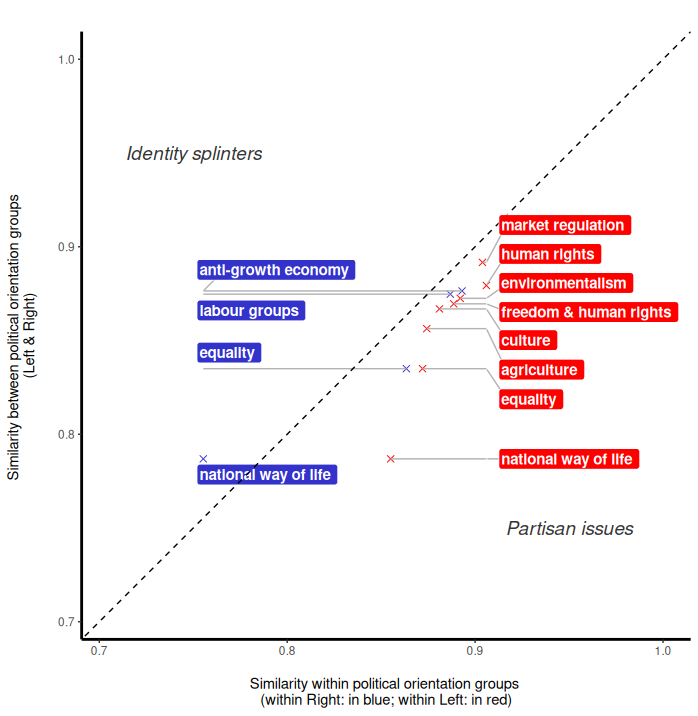}
    \caption{Scatterplot illustrating “partisan issues” and “identity splinters” at the EU level, for  the Left (in red) and the Right (in blue). Each point represents an issue for which the difference in mean similarity within the Left (or Right) group versus between the Left \& Right groups is statistically significant (p < 0.05). Following the framework proposed in Figure~\ref{fig:theory}.b., issues appearing below the x = y diagonal are classified as “partisan issues”, indicating greater similarity across countries within the Left (or Right) group compared to similarity between groups. Issues above the diagonal are classified as “identity splinters”, reflecting topics where intra-group similarity within the Left (or Right) is lower than similarity between the Left \& Right. Only issues with statistically significant within-group and between-group mean differences (at the 0.05 significance level) are shown. Tables containing the results of all the t-tests are available in the section S.8 of the Supplement}
    \label{fig:splinters_partisan}
\end{figure}

Following the conceptual framework presented in Subsection \ref{subsec:salience_framing}, Figure~\ref{fig:splinters_partisan} displays the issues for which there is a statistically significant difference in how similarly an issue is framed within (x-axis) and across (y-axis) political orientation groups, at the EU level. The figure shows that ten issues ---\textit{anti-growth economy}, \textit{environmentalism}, \textit{equality}, \textit{freedom \& human rights}, \textit{human rights}, \textit{labour groups}, \textit{market regulation}, \textit{culture}, \textit{minority groups}, and \textit{agriculture}--- are partisan issues. Only the \textit{equality} issue is to be considered partisan for both political orientation groups, meaning that the framing of this issue is specific to each group and clearly distinguishes them.
Other issues, such as \textit{national way of life}, \textit{environmentalism}, \textit{market regulation}, \textit{culture}, \textit{minority groups}, and \textit{agriculture}, are partisan only for Left-leaning groups. In contrast, \textit{anti-growth economy} and \textit{labour groups} are partisan only for the Right wing. These partisan issues are particularly relevant as topics of identity differentiation between the Left and Right at the EU level; we also hypothesize that they have the potential to be leveraged through propagandistic use of identity for deepening perceived divisions across the political spectrum, and mobilizing ideological groups, at the EU level. Figure~S.11 and Table~S.4 in the Supplement allow readers to visually explore partisan issue framing differences at the European level, and provide a detailed commentary on our findings for a subset of partisan issues of interest.
On the other hand,  Figure~\ref{fig:splinters_partisan} highlights that, for the Right side of the political spectrum, there is also one “identity splinter” issue, which is \textit{national way of life}, for which average similarity between political orientation groups is higher (at the 0.05 significance level) than that within (Right leaning) political orientation groups.
To examine the content of these issues and assess the extent to which discourse on these varies by political orientation (i.e., Left vs. Right), in section S.6 of the Supplement (see Figure~S.12, and Table~S.5) we analyse differences in framing between Left-wing and Right-wing users, at the EU level, for identity splinters, including those identifiable in Figure~\ref{fig:distance}, but for which these average similarity differences are not statistically significant at the 0.05 level.

\FloatBarrier

\subsection{Within-country issue framing similarity between Left- and Right-leaning user groups}\label{subsec:framing_country}

\begin{table}[!h]
\centering
\footnotesize
\setlength\extrarowheight{3pt}
\label{tab:country_sim}
\begin{tabularx}{\linewidth}{|p{1.4cm}|c|Y|Y|}
\hline
\textbf{Country} & \makecell[c]{\textbf{Average} \\ \textbf{Left-Right} \\ \textbf{similarity}} & \makecell[c]{\textbf{Top 3} \\ \textbf{issues by similarity}} & \makecell[c]{\textbf{Last 3} \\ \textbf{issues by similarity}} \\ \hline
Germany & 0.8866 & \makecell[c]{non-economic dem. groups (0.9278) \\ residual issue (0.9127) \\ culture (0.9101)} & \makecell[c]{equality (0.8258) \\ freedom \& human rights (0.8353) \\ democracy (0.8370)} \\
\hline
Spain & 0.8870 & \makecell[c]{non-economic dem. groups (0.9528) \\ minority groups (0.9459) \\ residual issue (0.9369)} & \makecell[c]{national way of life (0.7604) \\ internationalism (0.7987) \\ democracy (0.8133)} \\
\hline

Poland & 0.8971 & \makecell[c]{non-economic dem. groups (0.9528) \\ market regulation (0.9293) \\ minority groups (0.9252)} & \makecell[c]{national way of life (0.8062) \\ democracy (0.8376) \\ equality (0.8393)} \\
\hline
France & 0.9077 & \makecell[c]{non-economic dem. groups (0.9552) \\ minority groups (0.9455) \\ residual issue (0.9409)} & \makecell[c]{national way of life (0.8231) \\ equality (0.8640) \\ europe (0.8692)} \\
\hline

\makecell[l]{The\\ Netherlands} & 0.9084 & \makecell[c]{non-economic dem. groups (0.9493) \\ traditional morality (0.9426) \\ residual issue (0.9422)} & \makecell[c]{national way of life (0.8226) \\ democracy (0.8286) \\ agriculture (0.8651)} \\
\hline

Italy & 0.9254 & \makecell[c]{non-economic dem. groups (0.9549) \\ residual issue (0.9548) \\ labour groups (0.9458)} & \makecell[c]{democracy (0.8624) \\ decentralisation (0.8646) \\ national way of life (0.8699)} \\
\hline
Belgium & 0.9379 & \makecell[c]{non-economic dem. groups (0.9732) \\ minority groups (0.9607) \\ residual issue (0.9582)} & \makecell[c]{democracy (0.8726) \\ decentralisation (0.8979) \\ national way of life (0.9137)} \\
\hline
Slovenia & 0.9590 & \makecell[c]{non-economic demographic \\ groups (0.9769) \\ residual issue (0.9775) \\ human rights (0.9769)} & \makecell[c]{national way of life (0.8968) \\ traditional morality (0.9303) \\ democracy (0.9320)} \\
\hline
\end{tabularx}
\caption{Summary of the average issue similarities between the Left and the Right, by country, and list of three issues by country-specific similarity.}
\end{table}

The proposed approach also enables us to compare how issues are framed by Left- and Right-leaning groups within each country. Table~2 shows average Left-Right similarity scores by country, alongside the top three and bottom three issues where Left-Right alignment within the country is strongest and weakest, respectively.  Results indicate varying levels of similarity among political orientation groups, with average similarity scores ranging from 0.8867 (Germany) to 0.9590 (Slovenia). Notably, issues with the highest similarity scores consistently include non-economic demographic groups, suggesting that the framing of discussions around demographic aspects (e.g., being young, old, etc.) are largely shared across political divides. \\

Other high-similarity issues, albeit with some variation, include traditional morality and minority groups,  which seem to be framed in a similar way by Left- and Right-leaning users groups, especially in Spain, France, and Belgium. Conversely, the issues with the least similarity are more country-specific but generally include themes such as \textit{national way of life}, \textit{democracy}, and \textit{equality}, suggesting that discourse around these dimensions is likely more framed along ideological lines. For example, in Germany, the issues with the least alignment between the Left and Right are \textit{equality}, \textit{freedom \& human rights}, and \textit{democracy}. Similarly, in France and in the Netherlands, \textit{national way of life} and \textit{democracy} show substantial Left-Right heterogeneity. Belgium and Slovenia exhibit, overall, the highest similarity scores (0.9379 and 0.9590), indicating a closer alignment between Left and Right user groups within these countries. In contrast, Germany and Spain are near the bottom of the rankings (0.8866 and 0.8870), suggesting relatively more heterogeneity between political orientations on identity-related issues compared to other countries. For more details on within-country Left-Right similarities and differences in issue content, as well as a detailed explanation of the qualitative results and findings, see Supplement section S.7.\\

\begin{figure}
    \centering
    \includegraphics[width=0.75\linewidth]{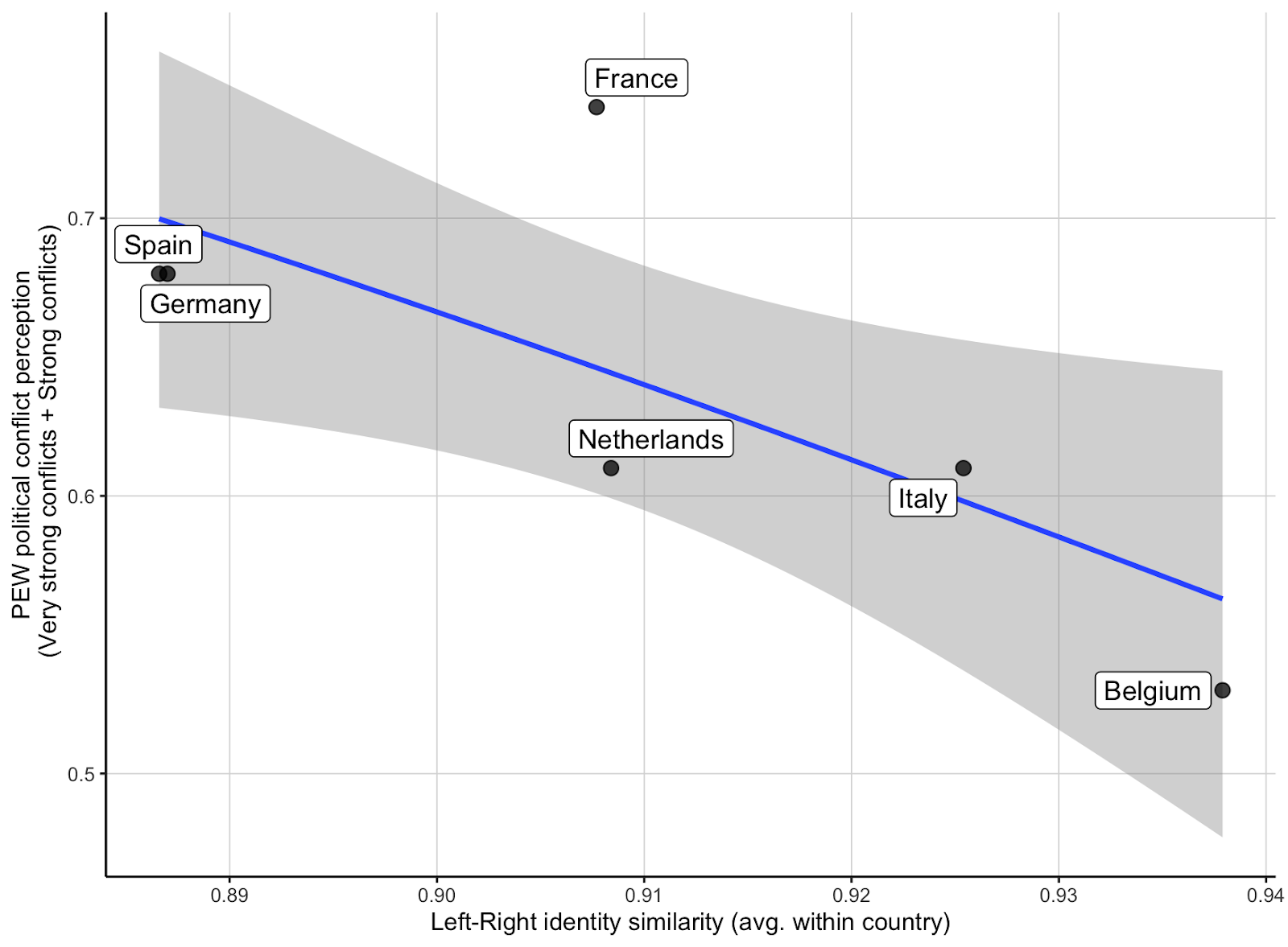}
    \caption{Scatterplot of PEW perceived political conflict score as a function of average similarities between the Left and the Right, by country. Line represents fitted values based on a GLM with a quasibinomial family, shaded area represents a 90\% Confidence Interval.}
    \label{fig:PEW}
\end{figure}

To explore the relevance of our Left-Right identity similarity metric as a measure of political identity fragmentation at the country level, we conduct a correlation analysis between our metric and an external measure of perceived political conflict. For this comparison, we used the Spring 2022 data from the PEW Global Attitudes and Trends survey, which assesses public perceptions of political conflict intensity by country. The survey includes a question on perceived political conflicts: "\textit{In your opinion, in [survey country], are the conflicts between people who support different political parties very strong, strong, not very strong, or are there no conflicts at all?}" Responses were categorized into five options (Very strong conflicts, Strong conflicts, Not very strong conflicts, No conflicts, Don’t know/Refused to answer).
The PEW 2022 survey covers eight (8) EU countries, therefore, our analysis is limited to the six countries that overlap between the PEW 2022 survey \citep{pew2022methodology, pew2022partisanconflict} and our study, that is: Belgium, France, Germany, Italy, the Netherlands, and Spain. 
Figure~\ref{fig:PEW} shows the positioning of these countries for the two metrics.

We calculate the correlation between our Left-Right identity similarity index and the proportion of survey respondents who selected either "Very strong conflicts" or "strong conflicts" in response to the question about political conflict intensity. We find a negative Pearson correlation coefficient (r = -0.7383) between our similarity metric and the percentage of respondents reporting "Very strong" or "strong" conflicts by country. Despite the weak power of this statistic due to the number of observations, this strong negative correlation suggests that higher average similarity between Left and Right within a country is associated with lower levels of perceived political conflict, while lower identity similarity correlates with higher perceived conflict. 
Interestingly, the country for which the fitted value of the political conflict index deviates the most from the observed one is France. France’s perceived political conflict intensity appears to be particularly high with respect to the observed Left-Right similarity metric. This finding is consistent with previous work based on expert surveys \citep{bakker2012complexity}, which shows that dimensions other than the Left–Right axis structure the political conflict in France and that political competition there requires at least two dimensions to explain the positioning of parties. Similarly, more recent work \citep{ramaciotti2022inferring} finds that political attitudes in the French Twittersphere are strongly fragmented with respect to sentiments toward elites. This suggests that, as indicated by the large positive deviation observed for France in this study, political conflict there goes beyond a simple Left–Right orientation.

These results provide preliminary evidence that the proposed approach, and in particular our Left-Right identity similarity metric, captures aspects of real-world societal fragmentation and perceived political identity-related  conflicts, as also reflected in the PEW survey.

\FloatBarrier
\section{Conclusion}
This study is the first to infer political-identity cues and discursive constructs directly from a very large sample of Social Media bios (more than 1.2 million) from eight countries, to analyse issue salience and framing, and to map the degree to which the expressed identities of user groups align within and across EU member states and political orientations.\\
The proposed approach allowed us to reveal important aspects of expressed identity heterogeneity across the political spectrum and among the investigated countries: First, allusions to \textit{national way of life},  \textit{democracy}, and \textit{decentralization} emerged as particularly divisive, showing considerable variation both within and between countries. These results, which align with H3 (and partially also with H1), are even more relevant since these three issues lie at the core of long-standing debates on European integration \citep{kohler2007debating}, and therefore also influence evolutionary pressures on European institutions, and related tensions between the national and the EU governance level.
Second, our approach allowed us to identify a set of “partisan issues” (\textit{anti-growth economy}, \textit{environmentalism}, \textit{equality}, \textit{freedom \& human rights}, \textit{human rights}, \textit{labour groups}, \textit{market regulation}, \textit{culture}, \textit{minority groups}, and \textit{agriculture}) that reflect major identity differences between the Left and Right, at the EU level. 
Issues such as \textit{environmentalism} and \textit{human rights} were, among others, identified as partisan issues associated with the Left. This finding, which is consistent with H2, partially aligns with other recent studies on the evolution of identity and values in Europe on the Left side of the political spectrum \citep{turovsky2025challenged}.\\
As suggested by \citet{horz2024identity},  these partisan identity-related issues and related norms are particularly important, as they can be exploited as propagandistic instruments by political elites seeking to mobilise ideologically aligned groups. More specifically, identity-focused propaganda may increase the salience of partisan identities, thereby also reshaping mobilisation incentives. Moreover, by raising the prominence of these issues in political discourse ---via media framing, campaign messaging, or public debate--- communicators, such as the political leaders referred to in Subsection \ref{subsec:framing}, can (intentionally or not) amplify perceptions of identity-based political polarization, further deepening divisions across the political spectrum, and strengthening (historical or new) identity cleavages \citep{zollinger2024cleavage} within and across countries. 
Our work therefore provides a methodologically sound approach and empirical grounding for future analyses, enabling us and other researchers to investigate online identity expression and identity-based political mobilization in greater depth. More specifically, in future work we aim to use this framework to examine whether, and to what extent, expressed identity dynamics and alignment help explain online political behaviour and mobilisation; for example, participation in petition campaigns and political crowdfunding initiatives.\\

In addition, our work highlights that there is another issue type particularly relevant to the study of political identity and its online expression, which we call “identity splinters”. We recall that these are issues for which the average similarity among EU countries within the Right (or Left) side of the political spectrum is lower than the average similarity between the Right and the Left. Overall, we observe higher issue-framing heterogeneity (i.e., lower average similarity) on the Right side of the political spectrum compared to the Left. Our findings hence provide stronger support for H3 on the Right than on the Left side of the political spectrum.
This category of issues, for which differences are statistically significant only for the issue \textit{national way of life}, shows greater average heterogeneity within the Right, at the EU level, than between Left and Right user groups. As also revealed by the qualitative analysis of national way of life framing specificities in France and Italy, issues that exhibit lower levels of alignment within the same ideological group across EU countries are a potential source of intra-ideological tensions. Observed issue-framing differences within the Right, at the European level, can, among other things, complicate users' identification with a (“unified”) Right‑wing identity, as the identified differences highlight national perspectives that resist simplification along traditional ideological lines.\\

Finally, within-country issue framing similarity metrics presented in Subsection \ref{subsec:framing_country}, show that Left-Right differences in expressed identities, at the country level, capture underlying social and political divides that align with public perceptions of conflict intensity. 
The proposed approach and the aforementioned findings also set the stage for future research to expand, across different geographic areas, attitudinal dimensions, and over time, at the desired granularity level, enabling through the proposed approach a richer grounded understanding of how differential issue-based self-identification correlates with shifts in political mobilization and identity-related ideological conflicts.

\paragraph{Acknowledgments}
The views and opinions expressed are solely those of the authors and do not necessarily reflect those of their institutions (INRIA, Sciences Po, CNRS, Ca’ Foscari, Lattice), of the European Union, or of the European Commission. Neither these institutions nor the European Union or the European Commission can be held responsible for them.\\
We would like to express our sincere gratitude to our colleagues at Sciences Po, at INRIA, and to other researchers who provided feedback on this article. In particular, we are deeply thankful to  Mathilde Emeriau, Ismail Harrando, Alex Kindel, Jan Rovny, and Brandon Stewart for their thoughtful comments and constructive feedback, which allowed us to greatly improve this work. 

\paragraph{Funding Statement}
Authors (C.S., J.C., P.R.) acknowledge funding from the European Union’s Horizon Europe programme under grant agreement ID 101094752: Social Media for Democracy (SoMe4Dem).
C.S. acknowledges funding from INRIA.
P.R. acknowledges funding from Project Liberty Institute project ``AI-Political Machines’’.
This project benefited from resources from the European Polarisation Observatory (EPO) of CIVICA Research (co-)funded by EU’s Horizon 2020 programme under grant agreement No 101017201 and from the Very Large Research Infrastructure (TGIR) Huma-Num of the CNRS, Aix-Marseille Université and Campus Condorcet.

\paragraph{Competing Interests}
The authors declare that they have no competing interests related to this work. There are no financial, personal, or professional relationships that could be perceived as influencing this research or its findings.

\paragraph{Data Availability Statement}
To comply with GDPR regulations and X/Twitter's terms of service, we have produced, and will make available on Harvard Dataverse, an anonymized version of the data and the R code used to estimate the STM model and generate the results and the figures presented in the main paper. The R code also allows the creation of many of the figures and tables from the supplementary materials, where possible without violating GDPR regulations and X/Twitter's terms of service.


\paragraph{Ethical Standards}
Informed consent was not required for this study, as the data collected from Twitter/X is publicly available through the platform’s interface. Only publicly accessible user bios were retrieved and included in the analysis. Bios from users with restricted privacy settings were not accessible and are therefore excluded by design from the data collection. 

\paragraph{Author Contributions}

Conceptualization: all (C.S., J.C., P.R.);
Methodology: all (C.S., J.C., P.R.);
Formal Analysis:  C.S.;
Investigation:  C.S. \& P.R.;
Data Curation: P.R.;
Software: C.S. \& P.R.;
Visualization: C.S.;
Writing – Original Draft Preparation:  C.S \& P.R.;
Writing – Review \& Editing: all (C.S., J.C., P.R.);
Funding Acquisition : P.R..
\newpage
\printbibliography






\end{document}